\documentclass{aa} % for a referee version
\usepackage{longtable}
\usepackage{supertabular}
\usepackage{graphics}
\usepackage{rotating}
\usepackage{natbib}
\begin{document}
\title{The AGN fraction -- velocity dispersion relation in clusters of galaxies} 
\author{Paola Popesso\inst{1} \& Andrea Biviano\inst{2}}
\institute{ European Southern Observatory, Karl Scharzschild Strasse 2, D-85748
\and INAF - Osservatorio Astronomico di Trieste, via G. B. Tiepolo 11, I-34143, Trieste, Italy}

\abstract {Some previous investigations have found that the fraction
($f_{AGN}$) of active galactic nuclei (AGNs) is lower in clusters than
in the field. This can result from the suppression of galaxy-galaxy
mergers in high-velocity dispersion ($\sigma_v$) clusters, if the
formation and/or fueling of AGNs is directly related to the merging
process.}{We investigate the existence of a relation between $f_{AGN}$
and $\sigma_v$ in galaxy clusters in order to shed light on the
formation and evolution processes of AGNs and cluster galaxies.}
{Using data from the Sloan Digital Sky Survey  we determine
$f_{AGN}$ and $\sigma_v$ for the clusters in two samples, extracted
from the catalogs of Popesso et al. (2006a) and Miller et al. (2005),
and excluding clusters with significant evidence for substructures.}
{We find a significant $f_{AGN}-\sigma_v$ anti-correlation. Clusters with $\sigma_v$ lower and, respectively, higher than 500 km
s$^{-1}$ have AGN fractions of $0.21 \pm 0.01$ and $0.15 \pm 0.01$, on
average. The $f_{AGN}-\sigma_v$ relation can be described by a model
that assumes $f_{AGN}$ is proportional to the galaxies merging rate,
plus a constant.}  {Since $f_{AGN}$ increases with decreasing
$\sigma_v$, AGNs are likely to have played a significant r\^ole in
heating the intra-cluster medium and driving galaxy evolution in
cluster precursors and groups.}


\authorrunning{Popesso \& Biviano}

\maketitle

\section{Introduction}
\label{s-intro}
Whether the fraction of active galactic nuclei (AGN hereafter) is
environment-dependent or not is a matter of debate. Claims that
clusters contain a lower fraction of AGNs ($f_{AGN}$ hereafter) than
the field (Gisler 1978; Dressler et al. 1985, 1999; Hill \& Oegerle
1993; Rines et al. 2005) are in disagreement with the recent finding
of Miller et al. (2003) that $f_{AGN}$ does not vary with the local
density. Such a discrepancy is explained by the finding of Kauffmann
et al. (2004) that high-luminosity AGNs do avoid high-density regions,
but low-luminosity AGNs do not; the latter dominate current AGN
samples, but were normally missed in older AGN surveys. The quoted
results were all based on optically-selected AGNs. X-ray observations
have found a general over-density of X-ray emitting AGNs relative to
field counts in the external regions of galaxy clusters (Henry \&
Briel 1991; Lazzati et al. 1998; Cappi et al. 2001; Martini et
al. 2006 and references therein). However, the observed over-densities
do not necessarily imply that the fraction of cluster galaxies hosting
an AGN is higher than the corresponding fraction of field galaxies.

A lower $f_{AGN}$ in clusters relative to the field could be explained
in terms of a decreased galaxy-galaxy merger efficiency in
clusters. In fact, AGNs are thought to be fueled, or even formed, by
the merger of gas-rich galaxies, which leads to rapid nuclear inflows
of gas (e.g. Barnes \& Hernquist 1992; Springel et al. 2005, S05
hereafter). The decreased merger efficiency in clusters arises from
the high relative velocities of cluster galaxies, since the merger
rate scales roughly as $\sigma_v^{-3}$ for high values of $\sigma_v$,
where $\sigma_v$ is the cluster velocity dispersion (Mamon 1992;
Makino \& Hut 1997). On the other hand, the low $\sigma_v$ and high
density of {\em groups} make them a very favorable environment for
galaxy-galaxy mergers, probably even more favorable than the field.
An indication that this might indeed be the case comes from the high
values of $f_{AGN}$ (typically $\ga 30$\%) reported for galaxy compact
groups (Coziol et al. 2000, 2004; Turner et al. 2001; Tovmassian et
al. 2006), and from the relatively high $f_{AGN}$ values found for
``open clusters'' in the early investigation of Gisler (1978).

In this letter the existence of a relation between cluster $f_{AGN}$
and $\sigma_v$ is investigated. For this purpose, we use two
samples of nearby (redshift $z < 0.08$) galaxy clusters with
data from the galaxy spectroscopic sample of the Sloan Digital Sky
Survey (SDSS). In Sect.~2 we briefly describe our cluster
samples, and how we determined $\sigma_v$ and $f_{AGN}$. In Sect.~3
we analyze the relation between these two quantities.  In Sect.~4 we
discuss our results and provide our conclusions. Throughout this
letter, we use $H_0=70$ km s$^{-1}$ Mpc$^{-1}$ in a flat cosmology
with $\Omega_0=0.3$, and $\Omega_{\Lambda}=0.7$.

\section{The cluster sample, $\sigma_v$, and $f_{AGN}$}
\label{s-data}
The optical data used in this paper are taken from the SDSS (see,
e.g., York et al. 2000; Abazajian et al. 2004). We use two
nearby cluster samples in our analysis, that of Popesso et
al. (2006a, PBBR hereafter), and the C4 cluster catalog of Miller
et al. (2005). Each cluster sample has its own merits. All clusters in
the PBBR have X-ray counterparts (although not all of them are X-ray
selected). The C4 catalog is entirely optically-selected, and has the
advantage of having a well understood selection function.

PBBR have selected the cluster members by the method of Katgert et
al. (2004), shown to be reliable by tests performed on clusters
extracted from cosmological simulations (van Haarlem et al. 1997;
Biviano et al. 2006). PBBR have determined virial masses, virial
radii ($r_{200}$s), and velocity dispersions ($\sigma_v$s) using the
cluster members with cluster-centric distances $\leq r_{200}$. We
refer the reader to PBBR for all the relevant details. The sample
of PBBR contains 137 clusters with at least 10 member galaxies within
$r_{200}$. Ten member galaxies are considered to be the minimum number
for a reliable determination of a cluster $\sigma_v$ (Girardi et
al. 1993; Zabludoff \& Mulchaey 1998). Using the same methods of PBBR
we select the cluster members in the C4 clusters, and then determine
the $\sigma_v$s of those clusters with at least 10 member galaxies
within their $r_{200}$s (609 clusters in total).

From both the PBBR and the C4 samples, we exclude those clusters with
significant evidence for subclustering (probability $<1$\%), according
to the test of Dressler \& Shectman (1988).  In this way we reduce the
occurrence of very wrong $\sigma_v$-estimates (see, e.g., van Haarlem
et al. 1997; Biviano et al. 2006).  There are 17 and, respectively,
115 clusters with subclustering in the PBBR and C4 samples.

We use all galaxies from the SDSS spectroscopic sample, with no
magnitude limit, to determine the cluster velocity dispersion, since
there is no evidence for segregation of cluster galaxies in velocity
space, apart for the very bright galaxies (Biviano et al.
1992). However, the fraction of AGN galaxies does depend on the galaxy
mass (Kauffmann et al. 2003), therefore we need to sample the cluster
galaxy populations down to the same absolute magnitude limit, that we
set to $M_r=-20.0$. We then select only those clusters that are
sampled deeper than this limit, i.e. those for which the limiting
magnitude $r=17.77$ of the SDSS spectroscopic survey corresponds to an
absolute magnitude $M_r \geq -20.0$, and with $>5$ members with
$M_r \leq -20.0$. Choosing a much brighter magnitude limit would leave
us with too few member galaxies per cluster, while choosing a much
fainter magnitude limit would leave us with too few clusters sampled
down to the chosen limit. Our final samples contain 63 (PBBR sample)
and 261 (C4 sample) clusters; all their $\sigma_v$-estimates are based
on $\geq 20$ galaxies.

\begin{figure}
\begin{center}
\begin{minipage}{0.40\textwidth}
\resizebox{\hsize}{!}{\includegraphics{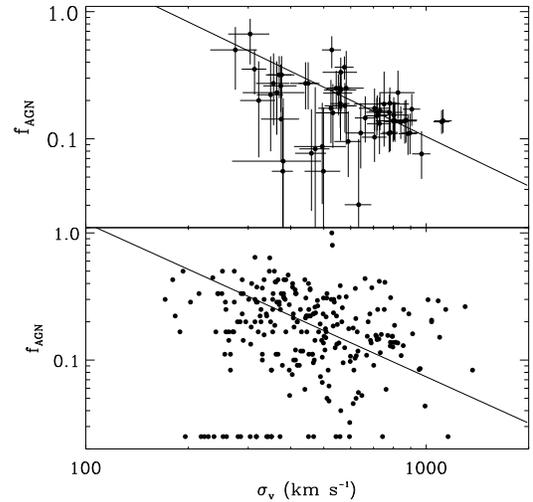}}
\end{minipage}
\end{center}
\caption{Top panel: The $f_{AGN}$--$\sigma_v$ diagram in
logarithmic scale, for the 63 clusters of the PBBR sample. The solid
line is the best-fit orthogonal linear regression in logarithmic
space. 1-$\sigma$ error-bars are determined following Gerhels (1986).
Bottom panel: same as panel a, for the 261 clusters of the C4
sample. Error-bars are not shown for the sake of clarity, but are
comparable to those shown in panel a. The 23 clusters with null values
of $f_{AGN}$ are plotted at $f_{AGN}=0.025$.}
\label{sigma_agn}
\end{figure}

The AGN classification of our cluster members is taken from Brinchmann
et al. (hereafter B04; see also Kauffmann et al. 2003), which identify
AGNs on the basis of the Baldwin et al. (1981, hereafter BPT)
criteria. BPT demonstrated that it is possible to distinguish AGNs
from normal star-forming galaxies by considering the intensity ratios
of two pairs of relatively strong emission lines, $y \equiv
\rm{[OIII]}\lambda 5007/\rm{H\beta}$ and $x \equiv
\rm{NII}\lambda6583/\rm{H\alpha}$. According to B04, all the galaxies
with $\log y \geq 0.61/(\log x - 0.47)+1.19$ (i.e. all objects above
the upper line in the BPT diagram shown in Fig.~1 of B04) are
classified AGNs. This classification is valid when the signal-to-noise
of all the four lines involved in the diagnostic is $\geq 3$. Galaxies
with a lower signal-to-noise of either the $\rm{[OIII]}\lambda 5007$,
or the $\rm{H\beta}$ line (or both) are classified AGNs if
$x>0.6$. AGN-classified galaxies have $> 40$\% of their $\rm{H\alpha}$
luminosity contributed by an AGN (B04).

We define $f_{AGN}$ as the fraction of AGNs among all the cluster
members within $r_{200}$ and with $M_r \leq -20.0$. Since
the completeness level is rather high for $M_r \leq -20.0$, and since
the sample we use is not biased in favor or against the selection of
emission-line galaxies, we deem it unnecessary to apply an
incompleteness correction to the derived values of $f_{AGN}$.

\section{Analysis and results}
\label{s-res}
The $f_{AGN}$ vs. $\sigma_v$ diagram is shown in
Figure~\ref{sigma_agn} for both cluster samples.  There is a
significant anti-correlation between the two quantities.  The
Spearman rank correlation coefficient (e.g. Press et al. 1986) is
$-0.45$ for the PBBR sample, and the null hypothesis of no correlation
is rejected with a probability of 0.9998.  The $f_{AGN}$--$\sigma_v$
anti-correlation is confirmed by the analysis of the C4 sample (the
Spearman correlation coefficient is $-0.16$, the no-correlation
hypothesis is rejected with a probability of 0.99).  Clusters with
$\sigma_v$ lower and, respectively, higher than 500~km~s$^{-1}$ have
AGN fractions of $0.21 \pm 0.03$ and $0.15 \pm 0.01$, on average, in
the PBBR sample (respectively $0.21 \pm 0.01$ and $0.16 \pm 0.01$, in
the C4 sample). We note that the $f_{AGN}$--$\sigma_v$
anti-correlation is not due to the $z$-dependence of the fraction of
low-luminosity AGNs in the SDSS spectroscopic sample (Kauffmann et
al. 2003) since there is no significant correlation between
$f_{AGN}$s and the average cluster redshifts in our samples.
This is expected since our clusters span a narrow $z$-range.

\begin{figure}
\begin{center}
\begin{minipage}{0.40\textwidth}
\resizebox{\hsize}{!}{\includegraphics{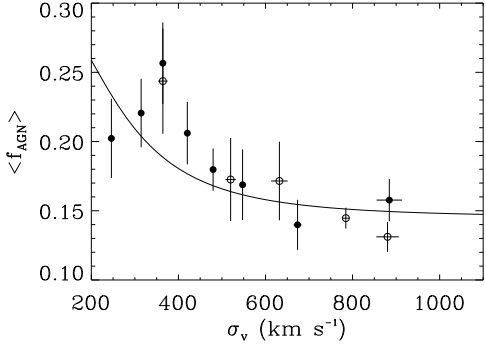}}
\end{minipage}
\end{center}
\caption{Biweight-average values of $f_{AGN}$ in several bins of
$\sigma_v$ for the PBBR (open symbols) and the C4 (filled symbols)
samples. The solid line displays the relation $f_{AGN} = 1.25 \, K(u)
+0.145$, where $K(u)$ is the merging rate function derived by Mamon
(1992), with $u=5.4 \sigma_0/(2 \, \sigma_v)$, and $\sigma_0=160$
km~s$^{-1}$.}
\label{mamonfit}
\end{figure}

We performed orthogonal linear regressions between $\log f_{AGN}$ and
$\log \sigma_v$, using the ODRPACK routine (Akritas \& Bershady
1996). For the C4 sample we obtain the best-fit relation $\log f_{AGN}
= (-1.21\pm0.12) \log \sigma_v + (2.50 \pm 0.24)$, which is consistent
(within the errors) with the relation obtained for the PBBR sample.
As an alternative description of the $f_{AGN}$-$\sigma_v$ relation we
also considered the expression $f_{AGN} = C \, K(u) + B$, where $u=5.4
\sigma_0/(2 \, \sigma_v)$, $\sigma_0$ is the galaxy internal velocity
dispersion, and $K(u)$ is the merging rate function derived by Mamon
(1992), which scales as $\sigma_v^{-3}$ for high, cluster-like values
of $\sigma_v$.  We take $\sigma_0=160 \, \rm{km~s}^{-1}$ as typical of
the $M_r \ga -20.0$ galaxies in our sample (we use the absolute
magnitude vs. $\sigma_0$ relation of Ziegler \& Bender 1997). Such a
relation, with $C=1.25$, $B=0.145$, provides an acceptable fit to the
biweight-average values of $f_{AGN}$ in several $\sigma_v$-bins both
for the PBBR ($\chi^2=6.0$ for 3 degrees of freedom, dof hereafter)
and the C4 sample ($\chi^2=10.2$ for 6 dof; see Figure~\ref{mamonfit};
see Beers et al.  1990 for the definition of the biweight-average
statistics).  On the other hand a constant $f_{AGN}$ model is rejected
(PBBR sample: $\chi^2=10.2$ for 4 dof; C4 sample: $\chi^2=18.4$ for 7
dof). We discuss the physical interpretation of this relation in
Sect.~\ref{s-disc}.

\begin{figure}
\begin{center}
\begin{minipage}{0.4\textwidth}
\resizebox{\hsize}{!}{\includegraphics{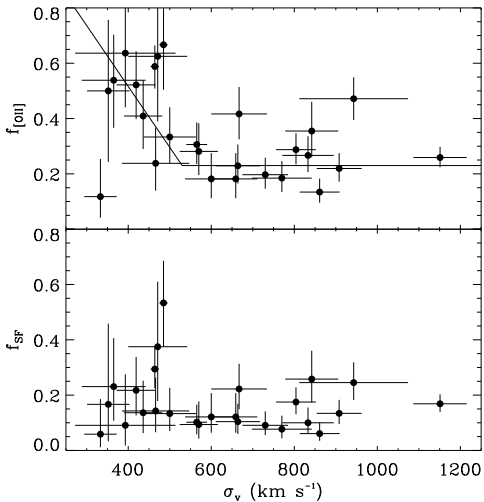}}
\end{minipage}
\end{center}
\caption{The relation between $f_{[OII]}$ (top panel) and $f_{SF}$
(bottom panel) and $\sigma_v$ for the sample of 26 clusters in
common between P06's and our samples. The solid line shows the
relation of P06 (see their eq.~(3)). 1-$\sigma$ error bars are shown.}
\label{sigma_oii}
\end{figure}

\section{Discussion \& conclusions}
\label{s-disc}
We have discovered a significant anti-correlation between $f_{AGN}$
and $\sigma_v$ in two samples of nearby galaxy clusters. Such an
anti-correlation is naturally expected if the formation and/or fueling
of AGNs is related to the galaxy-galaxy merging process (see, e.g.,
Veilleux et al. 2002; S05), since mergers of galaxies are impossible
if the galaxy relative velocities are high, as in rich galaxy
clusters.

The AGN fraction appears to be proportional to the merger rate in
galaxy systems (as determined theoretically by Mamon 1992), plus a
constant (see Figure~\ref{mamonfit}). The proportionality between
$f_{AGN}$ and the merger rate of cluster (or group) galaxies is
expected if galaxy-galaxy mergers increase the accretion rate of the
central black hole (and hence the AGN luminosity) over a time-scale
similar to the merger time-scale, as shown by the simulations of S05.

Why then is $f_{AGN}$ not zero for the highest-$\sigma_v$ clusters?
Even if the selected clusters in our sample do not show
significant evidence for subclustering, it is possible that most still
contain undetected subclusters, characterized by rather low,
group-like values of $\sigma_v$. Moreover, even if we have
selected cluster members in projected phase-space, this does not
eliminate completely the contamination by field galaxies, which
Biviano et al. (2006) estimate to be 17\%. Hence a substantial
fraction of the galaxies in our cluster samples could reside in
unidentified subclusters or in the field. If $f_{AGN}$ in subclusters
is as high as that in compact groups (Coziol et al. 2000, 2004; Turner
et al. 2001; Tovmassian et al. 2006) and if $f_{AGN}$ is higher in the
field than in clusters (see discussion below), this could explain why
the asymptotic value of $f_{AGN}$ in high-$\sigma_v$ clusters is not
zero.

Concerning the issue of whether $f_{AGN}$ in clusters is lower than
$f_{AGN}$ in the field, the result clearly depends on the selection of
the cluster sample. We find an average $f_{AGN}$ of $0.18 \pm 0.02$ for
our clusters, and this is value is consistent with the lower limit
obtained for the AGN fraction in the field, $f_{AGN}>0.20$ (Ho et
al. 1997; Carter et al. 2001; Kauffmann et al. 2003; Miller et
al. 2003; B04; Stasi\'nska et al. 2006). On the other hand, clusters
with $\sigma_v \geq 500$ km~s$^{-1}$ have an average $f_{AGN}$ of
$0.14 \pm 0.01$, significantly lower than the field value. Our cluster
AGN fraction is higher than previously reported values (Gisler 1978;
Dressler et al. 1985, 1999; Hill \& Oegerle 1993; Biviano et al. 1997;
Rines et al. 2005; Martini et al. 2006), but our estimate includes
low-luminosity AGNs which were probably missed in the previous
studies.

The $f_{AGN}$-$\sigma_v$ anti-correlation is not inconsistent
with the lack of a correlation between $f_{AGN}$ and the galaxy number
density (Miller et al. 2003).  We may in fact note that the merger
rate has an inverse cubic dependence on $\sigma_v$, but only a linear
dependence on the galaxy number density. If the AGN phenomenon is
indeed triggered by galaxy-galaxy mergers, the dependence of the AGN
fraction on the system $\sigma_v$ must be easier to detect than the
dependence on the density of the environment.

Our $f_{AGN}$-$\sigma_v$ plot of Figure~\ref{mamonfit} is very similar to
Fig.~6 in Poggianti et al. (2006, P06 hereafter), where the fraction
of cluster members with $\rm{EW(OII)} < -3$ ($f_{[OII]}$ hereafter) is
plotted as a function of the cluster $\sigma_v$ for a sample of 28
Abell clusters with SDSS spectroscopic data. The decreasing trend of
$f_{[OII]}$ with increasing $\sigma_v$ was interpreted by P06 in terms
of a decreasing fraction of star-forming galaxies with increasing
$\sigma_v$. Their conclusion assumes that the emission-line flux is
entirely stellar in origin. While it is true that AGN and
star-formation (SF hereafter) are probably related processes (e.g.
S05 and references therein) it is interesting to determine whether the
$f_{[OII]}$-$\sigma_v$ anti-correlation is driven mainly by the
suppression of SF or mainly by the suppression of AGNs in
high-$\sigma_v$ cluster galaxies.

There are 26 clusters in common between P06's sample and
ours. For these clusters we plot in Figure~\ref{sigma_oii} $f_{[OII]}$
vs. $\sigma_v$. Note that for the sake of comparison with P06, we have
used only the cluster members with $\rm{EW(OII)} < -3$.  The relation
of eq.~(3) in P06 is shown as a broken solid line.  Clearly we
reproduce their result. We then select only star-forming galaxies,
following the definition of B04, i.e.  excluding not only AGNs, but
also all the galaxies with a composite spectrum, i.e. those for which
up to 40\% of their H$\alpha$ luminosity might come from an AGN. The
fraction of star-forming galaxies, $f_{SF}$ vs. $\sigma_v$ is shown in
the same Figure. No trend is visible. The Spearman correlation
coefficient is $-0.05$ and the null hypothesis of no correlation has
an associated probability of $0.80$, hence cannot be rejected.
We then conclude that the $f_{[OII]}$ vs.  $\sigma_v$ is mostly, if
not entirely, due to the AGN contribution, i.e. the fraction of pure
star-forming galaxies does not depend on the cluster $\sigma_v$ (as
already found by Popesso et al. 2006b).

We conclude by pointing out that the $f_{AGN}-\sigma_v$ relation we
have found suggest that the AGN evolution must be linked to the
evolution of galaxy systems. The elevated merger rate in
low-$\sigma_v$ galaxy systems (groups or cluster progenitors) may
produce bursts of SF in their galaxies, followed by the formation of
AGNs.  AGN feedback may be crucial in quenching the SF in group
galaxies (Cooper et al. 2006), and in providing additional entropy to
the intra-group medium (e.g. Ettori et al. 2004). The hierarchical
growth of galaxy systems leads to an increase of their $\sigma_v$s,
which suppresses the merger processes and hence the formation of AGNs
(and associated feedback) in cluster galaxies.

\vspace{0.5cm}

We thank Vincenzo Mainieri and Paolo Tozzi for useful discussions.
We thank the anonymous referee for suggesting us to consider the
C4 cluster sample.

Funding for the SDSS and SDSS-II has been provided by the Alfred
P. Sloan Foundation, the Participating Institutions, the National
Science Foundation, the U.S. Department of Energy, the National
Aeronautics and Space Administration, the Japanese Monbukagakusho, the
Max Planck Society, and the Higher Education Funding Council for
England. The SDSS Web Site is http://www.sdss.org/. 
The SDSS is managed by the Astrophysical Research Consortium for the
Participating Institutions. The list of Participating Institutions can
be found at http://www.sdss.org/collaboration/credits.html/.

\end{document}